\documentclass[preprint2,times]{aastex6}
\usepackage{amsmath,amssymb,listings}
\usepackage{mathptmx}   %

\usepackage{etoolbox}

\makeatletter

\patchcmd{\NAT@citex}
  {\@citea\NAT@hyper@{%
     \NAT@nmfmt{\NAT@nm}%
     \hyper@natlinkbreak{\NAT@aysep\NAT@spacechar}{\@citeb\@extra@b@citeb}%
     \NAT@date}}
  {\@citea\NAT@nmfmt{\NAT@nm}%
   \NAT@aysep\NAT@spacechar\NAT@hyper@{\NAT@date}}{}{}

\patchcmd{\NAT@citex}
  {\@citea\NAT@hyper@{%
     \NAT@nmfmt{\NAT@nm}%
     \hyper@natlinkbreak{\NAT@spacechar\NAT@@open\if*#1*\else#1\NAT@spacechar\fi}%
       {\@citeb\@extra@b@citeb}%
     \NAT@date}}
  {\@citea\NAT@nmfmt{\NAT@nm}%
   \NAT@spacechar\NAT@@open\if*#1*\else#1\NAT@spacechar\fi\NAT@hyper@{\NAT@date}}
  {}{}

\makeatother

\newcommand{\mkmk}{{\texttt{Makemake}}}

\def\tauZeitskala{t}

\makeatletter
\let\jnl@style=\rm
\def\ref@jnl#1{{\jnl@style#1}}

\def\aj{\ref@jnl{AJ}}                   
\def\actaa{\ref@jnl{Acta Astron.}}      
\def\araa{\ref@jnl{ARA\&A}}             
\def\apj{\ref@jnl{ApJ}}                 
\def\apjl{\ref@jnl{ApJ}}                
\def\apjs{\ref@jnl{ApJS}}               
\def\ao{\ref@jnl{Appl.~Opt.}}           
\def\apss{\ref@jnl{Ap\&SS}}             
\def\aap{\ref@jnl{A\&A}}                
\def\aapr{\ref@jnl{A\&A~Rev.}}          
\def\aaps{\ref@jnl{A\&AS}}              
\def\azh{\ref@jnl{AZh}}                 
\def\baas{\ref@jnl{BAAS}}               
\def\bac{\ref@jnl{Bull. astr. Inst. Czechosl.}}
\def\caa{\ref@jnl{Chinese Astron. Astrophys.}}
\def\cjaa{\ref@jnl{Chinese J. Astron. Astrophys.}}
\def\icarus{\ref@jnl{Icarus}}           
\def\jcap{\ref@jnl{J. Cosmology Astropart. Phys.}}
\def\jrasc{\ref@jnl{JRASC}}             
\def\memras{\ref@jnl{MmRAS}}            
\def\mnras{\ref@jnl{MNRAS}}             
\def\na{\ref@jnl{New A}}                
\def\nar{\ref@jnl{New A Rev.}}          
\def\pra{\ref@jnl{Phys.~Rev.~A}}        
\def\prb{\ref@jnl{Phys.~Rev.~B}}        
\def\prc{\ref@jnl{Phys.~Rev.~C}}        
\def\prd{\ref@jnl{Phys.~Rev.~D}}        
\def\pre{\ref@jnl{Phys.~Rev.~E}}        
\def\prl{\ref@jnl{Phys.~Rev.~Lett.}}    
\def\pasa{\ref@jnl{PASA}}               
\def\pasp{\ref@jnl{PASP}}               
\def\pasj{\ref@jnl{PASJ}}               
\def\rmxaa{\ref@jnl{Rev. Mexicana Astron. Astrofis.}}%
\def\qjras{\ref@jnl{QJRAS}}             
\def\skytel{\ref@jnl{S\&T}}             
\def\solphys{\ref@jnl{Sol.~Phys.}}      
\def\sovast{\ref@jnl{Soviet~Ast.}}      
\def\ssr{\ref@jnl{Space~Sci.~Rev.}}     
\def\zap{\ref@jnl{ZAp}}                 
\def\nat{\ref@jnl{Nature}}              
\def\iaucirc{\ref@jnl{IAU~Circ.}}       
\def\aplett{\ref@jnl{Astrophys.~Lett.}} 
\def\apspr{\ref@jnl{Astrophys.~Space~Phys.~Res.}}
\def\bain{\ref@jnl{Bull.~Astron.~Inst.~Netherlands}} 
\def\fcp{\ref@jnl{Fund.~Cosmic~Phys.}}  
\def\gca{\ref@jnl{Geochim.~Cosmochim.~Acta}}   
\def\grl{\ref@jnl{Geophys.~Res.~Lett.}} 
\def\jcp{\ref@jnl{J.~Chem.~Phys.}}      
\def\jgr{\ref@jnl{J.~Geophys.~Res.}}    
\def\jqsrt{\ref@jnl{J.~Quant.~Spec.~Radiat.~Transf.}}
\def\memsai{\ref@jnl{Mem.~Soc.~Astron.~Italiana}}
\def\nphysa{\ref@jnl{Nucl.~Phys.~A}}   
\def\physrep{\ref@jnl{Phys.~Rep.}}   
\def\physscr{\ref@jnl{Phys.~Scr}}   
\def\planss{\ref@jnl{Planet.~Space~Sci.}}   
\def\procspie{\ref@jnl{Proc.~SPIE}}   

\def\ptp{\ref@jnl{Prog.~Th.~Phys.}}   

\makeatother

\defcitealias{goli04}{G04}
\defcitealias{marl07}{M07}
\defcitealias{scvh}{SCvH}
\defcitealias{bl94}{BL94}
\defcitealias{mc14}{MC14}
\defcitealias{ensman94}{E94}

\newcommand{\Vekt}[1]{\mathbf{#1}}  

\newcommand{\dd}{{\rm d}}

\def\upartial{\partial}

\usepackage{grffile}

\newcommand{\K}[1]{}

\definecolor{Hellgrau}{gray}{0.7}

\newcommand{\MJ}{{M_{\textnormal{J}}}}
\newcommand{\RJ}{{R_{\textnormal{J}}}}

\newcommand{\ME}{{M_{\oplus}}}

\newcommand{\LSonne}{{L_{\odot}}}

\newcommand{\sigSB}{{\sigma}}

\newcommand{\mH}{{m_{\textnormal{H}}}}

\newcommand{\kB}{{k_{\textnormal{B}}}}

\newcommand{\XHI}{{X_{\textnormal{H}}}}

\newcommand{\delad}{{\nabla_{\!\textnormal{ad}}}}
\newcommand{\delrad}{{\nabla_{\!\textnormal{rad}}}}
\newcommand{\deltat}{{\nabla_{\!\textnormal{act}}}}  

\newcommand{\MP}{{M_{\textnormal{p}}}}
\newcommand{\RP}{{R_{\textnormal{p}}}}

\newcommand{\LP}{{L_{\textnormal{p}}}}
\def\Lint{\LP}

\newcommand{\RAkk}{{R_{\textnormal{acc}}}}

\newcommand{\RHill}{{R_{\textnormal{Hill}}}}
\newcommand{\RBondi}{{R_{\textnormal{Bondi}}}}
\newcommand{\kLiss}{{k_{\textnormal{Lissauer}}}}
\newcommand{\MPunkt}{{\dot{M}}}

\newcommand{\TNeb}{{T_{\textnormal{neb}}}}

\newcommand{\Mach}{{\mathcal{M}}}

\newcommand{\TSch}{{T_{\textnormal{s}}}}

\newcommand{\MStern}{{M_{*}}}

\newcommand{\cP}{{c_{\textnormal{p}}}}
\newcommand{\cV}{{c_{\textnormal{v}}}}

\newcommand{\kapR}{{\kappa_{\textnormal{R}}}}

\newcommand{\TZerst}{{T_{\textnormal{dest}}}}

\newcommand{\vFf}{{v_{\textnormal{ff}}}}
\newcommand{\rhoFf}{{\varrho_{\textnormal{ff}}}}

\newcommand{\ttherm}{{\tauZeitskala_{\textnormal{therm}}}}
\newcommand{\tAkk}{{\tauZeitskala_{\rm acc}}}

\newcommand{\fred}{{f_{\textnormal{red}}}}

\newcommand{\ceff}{{c_{\textnormal{eff}}}}

\newcommand{\mfWPh}{{\lambda_{\rm phot}}}
\newcommand{\HErad}{{H_{\Erad}}}

\newcommand{\Eint}{{E_{\textnormal{int}}}}
\newcommand{\Ekin}{{E_{\textnormal{kin}}}}

\newcommand{\Etot}{{E_{\textnormal{tot}}}}
\newcommand{\Erad}{{E_{\textnormal{rad}}}}

\def\Lkin{\LAkkmax}

\newcommand{\eint}{{e_{\textnormal{int}}}}
\newcommand{\ekin}{{e_{\textnormal{kin}}}}
\newcommand{\etot}{{e_{\textnormal{tot}}}}

\newcommand{\DF}[1][]{{D_{\textnormal{F}#1}}}

\newcommand{\rmin}{{r_{\rm min}}}
\newcommand{\rmax}{{r_{\rm max}}}

\newcommand{\rSchock}{{r_{\rm shock}}}
\newcommand{\LAkk}{{L_{\rm acc}}}

\newcommand{\LAkkmax}{{L_{\rm acc,~max}}}
\newcommand{\TSchock}{{T_{\rm shock}}}

\newcommand{\Llinks}{{L_{\rm downstr}}}

\newcommand{\Frad}{{F_{\rm rad}}}

\def\vSch{{ v_{\rm shock} }}
\def\etaklassisch{ \eta^{\rm kin} }
\def\etaphys{      \eta^{\rm phys} }

\def\Pram{P_{\rm ram}}

\newcommand{\cs}{{c_{\rm s}}}

\newcommand{\EPkt}{{\dot{E}}}

\newcommand{\Ae}[1]{{#1}}
\newcommand{\Aeneu}[1]{{#1}}

\begin{document}

\title[Planet formation accretion shock: Framework]{The \Ae{planetary accretion shock}:\\I.~Framework for radiation-hydrodynamical simulations and first results}
\author{Gabriel-Dominique Marleau\altaffilmark{1,2},
Hubert Klahr\altaffilmark{2}, Rolf Kuiper\altaffilmark{3}, Christoph Mordasini\altaffilmark{1}}
\affil{
\altaffilmark{1}{Physikalisches Institut, Universit\"{a}t Bern, Sidlerstr.~5, 3012 Bern, Switzerland}\\
\altaffilmark{2}{Max-Planck-Institut f\"ur Astronomie, K\"onigstuhl 17, 69117 Heidelberg, Germany}\\
\altaffilmark{3}{Institute for Astronomy and Astrophysics, Eberhard Karls Universit\"at T\"ubingen, Auf der Morgenstelle 10, 72076 T\"ubingen, Germany}
}
\email{gabriel.marleau@space.unibe.ch}

\begin{abstract}
The key aspect determining the post-formation luminosity of gas giants has long been considered to be the energetics of the accretion shock at the planetary surface. We use one-dimensional radiation-\hspace{0pt}hydrodynamical simulations to study the radiative loss efficiency and to obtain post-shock temperatures and pressures and thus entropies. The efficiency is defined as the fraction of the total incoming energy flux which escapes the system (roughly the Hill sphere), taking into account the energy recycling which occurs ahead of the shock in a radiative precursor. We focus here on a constant equation of state to isolate the shock physics but use constant and tabulated opacities. While robust quantitative results will require a self-consistent treatment including hydrogen dissocation and ionization, the results presented here show the correct qualitative behavior and can be understood semi-analytically. The shock is found to be isothermal and supercritical for a range of conditions relevant to core accretion (CA), with Mach numbers $\Mach\gtrsim3$. Across the shock, the entropy decreases significantly, by a few entropy units ($\kB/\mbox{baryon}$). While nearly 100 percent of the incoming kinetic energy is converted to radiation locally, the efficiencies are found to be as low as roughly 40 percent, implying that a meaningful fraction of the total accretion energy is brought into the planet. For realistic parameter combinations in the CA scenario, a non-zero fraction of the luminosity always escapes the system. This luminosity could explain, at least in part, recent observations in the young LkCa 15 and HD 100546 systems.
\end{abstract}


\section{Introduction}

Starting with the discovery of planetary and low-mass companions to 2M 1207, 1RXS 1609, HR 8799, and $\beta$ Pic
in the last decade \citep{chauvin04,lafreniere08,marois08,lagrange09},
photometric and spectroscopic direct observations of several dozen young ($\lesssim20$--100-Myr-old) objects have challenged and enriched  %
our knowledge about exoplanets, providing access to their theirmodynamic state,
chemically complex atmospheres, and otherwise unobtainable information
on the outer ($\gtrsim20$~au) architecture of planetary systems.
One major limitation, however, has been the difficulty of determining the masses of these objects,
which is of particular importance in the light of recent or upcoming surveys expected to detect several young objects
(e.g., LEECH, SPHERE, GPI, Project 1640, CHARIS; see \citealt{skemer14a,zurlo14,macintosh14,opp12,peterslimbach13} and references therein)
as they seek to provide constraints on the mass distribution \Ae{of planetary or very-low-mass companions}
\citep[e.g.][]{biller13,brandt14,clantongaudi15}.

While the uncertainty on the age of the parent star often remains considerable, it is, to first order, presumably random.
However, the conversion of a luminosity to a mass entails a theoretical, probably \textit{systematic} uncertainty:
that of the luminosity of a planet or low-mass object at the end of its formation,
as it enters into the evolutionary `cooling' phase\footnote{Deuterium burning might slow down the cooling but the argument remains the same.}.
This is a major source of uncertainty \citep{bowler16}.
Indeed, at those young ages, cooling
has not yet erased traces of the formation process,
as reflected in a planet's luminosity and radius (and thus also spectrum);
this happens on the Kelvin--Helmholtz timescale $t_{\rm KH}\equiv G\MP^2/RL\sim10^7$--$10^9$~yr.
Formation models up to now \citep[e.g.,][]{marl07,morda12_I} have only made predictions
in the limiting cases of `hot' and `cold starts', as discussed below,
without \Ae{however attempting to model the shock in detail}.  %

What is thought to be the key process setting the entropy of the gas
is the accretion shock in the runaway gas accretion phase \citep{marl07,spiegel12}.
This accretion shock is traditionally associated with core accretion
but it might also occur in some circumstances in the context of gravitational instability
(see the discussion in Section~8.1 of \citealp{morda12_I}).
When the planet becomes massive enough, it detaches from the local disk
and gas falls freely onto it.
The question is usually put in terms of what happens to the kinetic energy of the gas,
namely whether it is radiated away at the shock
or whether it gets added as thermal energy to the planet.
The extreme outcome of full radiative loss leads to `cold starts',
while the limiting case of no radiative loss leads to `hot starts',
as the resulting planets are then respectively colder or hotter \citep{marl07}.

\citet{morda13} found that the mass of the solid core (in the core-accretion framework)
correlates with post-formation luminosity but, as explained there,
this is due to a self-amplifying process based on the shock.
Thus the shock (or at least its computational treatment in formation calculations)
is crucial is setting the post-formation radius and luminosity.

It has been shown \citep{mc14} how to place joint constraints on the mass and initial entropy of an object
from a luminosity and age measurement.
Now, however, we take a first step towards \textit{predicting} this initial entropy
by presenting simulations of the shock efficiency, considering snapshots of the formation process.

In this first paper, we focus on the physics at the accretion shock
and the upstream region. Since ionization and dissociation act as energy sinks \citep{zeldovich67},
we focus on an ideal-gas equation of state (EOS) with constant heat capacity
and mean molecular weight to isolate the shock physics from the microphysics.
However, we use both constant and more realistic opacities.
Also to simplify the analysis, we assume here that the gas and the radiation couple perfectly
and therefore use `one-temperature' (1-$T$) radiation transport (discussed below).
A forthcoming paper will be concerned with the effects of dissociation and ionization  %
and will also address the importance of 2-$T$ radiation transport.
Finally, a subsequent work will also discuss how the shock results can be used in formation calculations
and perform this coupling.
Only with this will it be possible to predict directly %
post-formation entropies and thus the luminosities and radii of young gas giants.

\section{Model}
 \label{Theil:Modell}

\subsection{Physical picture}

Each simulation is meant as a snapshot of the accretion process
when the planet \Ae{is at} a radius $\RP=\rSchock$, the shock radius.  %
To follow gas accretion onto a growing planet which is detached from the nebula,
we let our simulation box extend from the top-most layers of the planet
to a large fraction of its accretion radius
$\RAkk$, defined through \citep{boden00}
\begin{equation}
\label{Gl:RAkk}
 \frac{1}{\RAkk} = \frac{1}{\kLiss\RHill} + \frac{1}{\RBondi},
\end{equation}
where
\begin{equation}
\RHill = a \left(\frac{\MP}{3\MStern}\right)^{1/3},\;\;\;\RBondi = \frac{G\MP}{{c_\infty}^2}
\end{equation}
are the Hill and Bondi radii, respectively,
with $a$ the semi-major axis of the planet of mass $\MP$ around a star of mass $\MStern$
and $c_\infty$ the sound speed in the disk at the planet's location.
The factor $\kLiss=1/3$ accounts for the findings of \citet{lissauer09}
that only the inner part of the material in the planet's Roche lobe is bound to it,
most of the gas in the volume flowing with the material in the disk \citep{morda12_I}.
The sphere of radius $\RAkk$ is thus the approximate region where gas should become bound to the planet,
both in terms of gravitational force compared to that from the star ($\RHill$)
and thermal energy compared to the planet's potential energy ($\RBondi$).
In the runaway phase, the $\RHill$ term usually (though marginally) dominates.

While global and local disk simulations have shown that the accretion
onto the protoplanet is highly three-dimensional \citep{ab09a,tanigawa12,dangeloboden13,ormel15b,fung15,szul16}  %
and possibly affected by magnetic fields in the gap and protoplanetary disk \citep[e.g.,][]{uribe13,keith15},
we take a first step here by using a spherically-symmetric set-up and neglecting magnetic fields.
This allows us to model in detail the last stages of the accretion process
on small scales around the proto-planetary object ($r \lesssim30~\RJ$).
This stages should remain similar in more complex geometries.

\K{   %
Each simulation is meant as a snapshot of the accretion process
when the planet has grown to $\RP=\rSchock$.
This is possible because %
the thermal timescale $\ttherm=\cP T \Delta M/L$ of the gas of mass $\Delta M$ in the simulation domain 
is much shorter than the accretion timescale $\tAkk=\Delta M/\MPunkt$. %
Thus the gas present can adjust to the energy input
}

Note that,
in the detached runaway phase, the continued accretion of solids (dust and planetesimals) by the planet
is important for setting the final mass of the core \citep{morda13}.
However, this accretion rate of solids is several orders of magnitude smaller
than that of gas and is therefore neglected here.

\subsection{Methods}
 \label{Theil:Kode}

For our one-temperature radiation hydrodynamics simulations, %
we use the static-grid version of the modular \mbox{(magneto-)}hy\-drodynamics code \texttt{PLUTO} (version 3; \citealp{mignone07,mignone12})
combined with the grey, 1-$T$ flux-limited diffusion (FLD) radiation transport  %
module \mkmk{} 
described and tested in \citet{kuiper10} and \citet{kuiperklessen13}, without ray tracing. %
We use the HLL hydrodynamical solver and the flux limiter $\lambda$ from \citet{lever81} given by
\begin{equation}
\label{Gl:lambdaFLD}
 \lambda(R) = \frac{2+R}{6+3R+R^2},
\end{equation}
where the radiation parameter $R$ is defined through
\begin{subequations}
\begin{align}
 \Frad &= -\DF \Vekt{\nabla}\Erad,\label{Gl:FStrahl}\\
 \DF &\equiv \frac{\lambda(R) c}{\kapR\varrho},\\
 R &\equiv \frac{\| \Vekt{\nabla}\ln \Erad \|}{\kapR\,\varrho},\label{Gl:RDef}  %
\end{align}
\end{subequations}
where $\Frad$ and $\Erad$ are the radiation flux and energy density, respectively,
$\kapR$ the Rosseland mean opacity,
$\rho$ the density, and $c$ the speed of light.
There is some freedom in the choice of the flux limiter's functional form but it is required to behave
asymptotically as \citep{lever84}
\begin{equation}
\lambda(R) \rightarrow \begin{cases}
 \frac{1}{3}, & R\ll1~~~\textnormal{(diffusion limit)}\\
 \frac{1}{R}, & R\gg1~~~\textnormal{(free-streaming limit)}
 \end{cases}
\end{equation}
to recover the limiting regimes of pure diffusion, where
$\Frad=\frac{1}{3}c\nabla\Erad/\kapR\rho$,
and free-streaming, where $\Frad=c\Erad$ in the direction opposite to the $\Erad$ gradient.

The local radiation quantity $R(\varrho,T,\Erad)$ defined in Equation~(\ref{Gl:RDef}) compares
the photon mean free path $\mfWPh=1/\kappa\varrho$
to the `radiation energy density scale height' $\HErad=\Erad/(\partial\Erad/\partial r)$;
in spherical coordinates it is given by
\begin{subequations}
\label{Gl:Rpraktisch}
\begin{align}
 R &= \frac{1}{\kappa\varrho\Erad} \left|\frac{\partial\Erad}{\partial r}\right| = \left|\frac{\partial\ln\Erad}{\partial\tau}\right|\\
   &= \frac{\mfWPh}{\HErad}.
\end{align}
\end{subequations}
Large $R$ values mean that the radiation energy density---and thus,
in the 1-$T$ approximation, the temperature---changes
over a shorter distance than photons get absorbed and re-emitted.

\subsection{Set-up}

We use a semi-open box fixed at some height in the atmosphere of the planet,
with a closed left, inner edge (towards the centre of the planet) at $r=\rmin$,
and start with an atmosphere of some arbitrary small height (e.g., 0.5~$\RJ$),
onto which gas falls from the outer edge of the grid at $\rmax$.
For the initial set-up, we calculate an atmosphere in hydrostatic equilibrium
with a constant luminosity $\LP=10^{-3}~\LSonne$ using the usual equations of stellar structure
(but Equation~(\ref{Gl:HydrostatStrukt_dLdr0}) as appropriate for an atmosphere):
\begin{subequations}
\label{Gl:HydrostatStrukt}
\begin{align}
\frac{\dd m_r}{\dd r} &= 4\pi r^2 \varrho, \\
\frac{\dd T}{\dd r} &= \deltat \frac{T}{P} \frac{\dd P}{\dd r},\\
\frac{\dd P}{\dd r} &= -\varrho \frac{G m_r}{r^2},\\
\frac{\dd L}{\dd r} &= \frac{\dd m_r}{\dd r}\left( \varepsilon - T \frac{\dd S}{\dd r}\right),\notag\\
                    &= 0,\label{Gl:HydrostatStrukt_dLdr0} %
\end{align}
\end{subequations}
where
$m_r$ is the mass interior to $r$ (dominated by $\MP$),
$P$, $T$, and $S$ are respectively the pressure, the temperature, and the entropy per mass,
$L=4\pi r^2\Frad$ is the luminosity,
$G$ the universal gravitational constant,
and $\varepsilon$ the energy generation rate.
The actual, adiabatic, and radiative gradients are given respectively by
\begin{subequations}
\begin{align}
 \deltat &= \min(\delad,\delrad)\label{Gl:Schwarzschild}\\
 \delad &= \frac{\gamma-1}{\gamma},\\
 \delrad &= \frac{3 L P \kappa}{64\pi\sigSB G m_r T^4}.
\end{align}
\end{subequations}
Equation~(\ref{Gl:Schwarzschild}) is the Schwarzschild criterion.
\Ae{(Note that convection therefore plays a role only in the initial profile;
in the radiation-\hspace{0pt}hydrodynamical simulations proper there is no convection
because of the assumption of spherical symmetry.)}
We use an adaptive step size for the integration
to resolve accurately the pressure and temperature gradients. %
This atmosphere is then smoothly joined onto a calculated accretion flow
for $\varrho$ and $v$. (See Equations~(\ref{Gl:vFf}ff) below.)
The goal of these efforts is (i)~to provide a numerically sufficiently smooth initial profile
while~(ii) beginning with a certain atmospheric mass to speed up the computation.

The grid is uniform from $\rmin$ to $\rmin+\Delta r$
\Ae{and has a high resolution} to resolve
sufficiently well the pressure gradient in the innermost part,
using by default $\Delta r=0.5~\RJ$ and $N=500$~cells there.
The other grid patch is a stretched segment out to $\rmax$,
with usually also $N=500$, i.e., a much smaller resolution.
This has proven to be stable and accurate.

\Ae{As gas is added to the simulation domain, quasi-hydrostatic equilibrium
establishes below the shock.
The shock position $\rSchock$ defining the top of the planet's atmosphere is simply given by
the location where the gas pressure is equal to the ram pressure.
The shocks moves in time as gas is added (inward or outward depending on the simulation),
usually at a negligible speed, i.e., $d\rSchock/dt\ll \vSch$,
where $\vSch$ is the pre-shock velocity.
Nevertheless, we always take this term into account when calculating mass or energy fluxes;
this possibly leads to slightly non-nominal effective accretion rates
but allows for a more accurate verification of energy conservation.}
We consider only data from after an early adjustment phase,
once the \Ae{lab-frame} accretion rate at the shock is equal
to the one set through the outer boundary conditions, described below.

\subsection{Boundary conditions}

For the hydrodynamics, 
reflective (zero-gradient) boundary conditions are used at $\rmin$ in the density, pressure, and velocity,
i.e., %
\begin{equation}
   \frac{{\rm d}P}{{\rm d}r} = \frac{{\rm d}\rho}{{\rm d}r} = \frac{{\rm d}v}{{\rm d}r}=0
\end{equation}

Since the condition ${\rm d}v/{\rm d}r=0$
ensures that no mass flows over the boundary,
it is not necessary to enforce hydrostatic equilibrium at $\rmin$.
In the radiation transport also, we prevent the flow of energy over $\rmin$ by using
\begin{equation}
 \frac{{\rm d}\Erad}{{\rm d}r} = 0.
\end{equation}

The outer edge of the grid $\rmax$ is set well outside of the atmosphere and away from the shock.
For the hydrodynamics, we choose an accretion rate and approximate the velocity
as the free-fall velocity:
\begin{equation}
\label{Gl:vFf}
   v(r) = \vFf(\rmax) = \sqrt{2G\MP\left(\frac{1}{\rmax} - \frac{1}{\RAkk}\right)},
\end{equation}
with $\RAkk$ defined in Equation~(\ref{Gl:RAkk}).
Mass conservation then determines the `free-fall density':
\begin{equation}
\label{Gl:rhoFf}
   \rhoFf(\rmax) = \frac{\MPunkt}{4\pi r^2 |v(\rmax)|}.
\end{equation}
The pressure gradient here too is required to vanish:
\begin{equation}
 \left.\frac{{\rm d}P}{{\rm d}r}\right|_\rmax = 0.
\end{equation}
We considered for some simulations a Dirichlet boundary condition
with $P=P(\rhoFf(\rmax),$ $\TNeb)$ for a nebula temperature $\TNeb$,
taken as $\TNeb=150$~K \citep[e.g.,][]{miz80}. %
This did not change the results significantly.

Finally, the radiation outer boundary condition is usually set to the flux-divergence-free condition
\begin{equation}
 \label{Gl:SRB_lambdaklein}
 \frac{\upartial r^2\Erad}{\upartial r}=0,
\end{equation}
which corresponds to a constant luminosity if the reduced flux $\fred$, defined in Section~\ref{Theil:Deltatau,fred},
is sufficiently close to 1.
However, even when the flux at the outer edge is rather in the diffusion regime,
we obtain similar results for a simple Dirichlet boundary condition on the radiation temperature.

\subsection{Microphysics}

To isolate the shock physics, we consider in this work a constant equation of state (EOS).
The EOS enters into the radiation-\hspace{0pt}hydrodynamical simulations through
the effective heat capacity ratio $\gamma\equiv \cP/\cV = P/\Eint+1$,
where $\Eint$ is the internal energy \Ae{per volume}, and through the mean molecular weight $\mu$.
Estimates presented in Appendix~\ref{Anh:Schockabschaetzung} suggest that
the hydrogen will be in molecular or atomic form at the shock.
Accordingly, $\gamma=1.44$ and $\gamma=1.1$ bracket the expected range,
while $\mu$ varies from $2.353$ to $1.23$ (see Figure~\ref{Abb:ungefSchPopSynth_kalt}).

\begin{figure*}
 \epsscale{0.6}
 \plotone{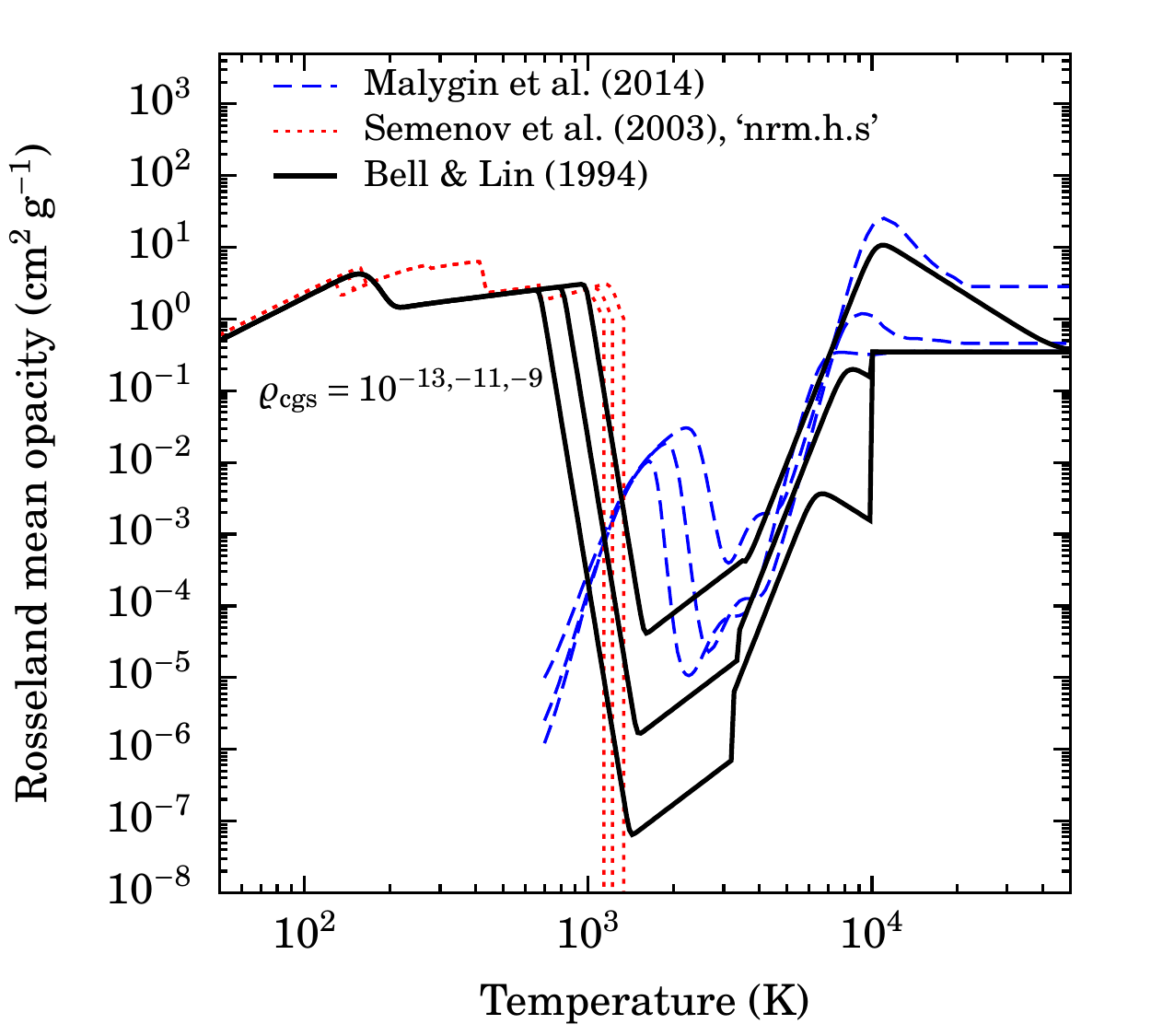}
\caption{
Gas, dust, and total Rosseland mean opacities from \citet{malygin14}, \citetalias{bl94},
and \citet{semenov03}. %
Three densities are shown: $\varrho=10^{-13,-11,-9}$~g\,cm$^{-3}$.
For the \citet{semenov03} opacities, we use their `nrm.h.s' model,
with dust grains made of `normal silicates' ([Fe/(Fe+Mg$)]=0.4$) as homegeneous spheres.
The \citet{malygin14} opacities are kept constant above the table limit
of $T=2\times10^4$~K.
}
\label{Abb:kappaSchnitt}
\end{figure*}

We consider both constant and tabulated opacities.
The contribution of the dust to the opacity dominates below approximately 1400--1600~K,
at which temperature the refractory components (olivine, silicates, pyroxene, etc.) evaporate \citep{poll94,semenov03}.
The standard opacity tables we use are the smoothed \citet[][hereafter~\citetalias{bl94}]{bl94} tables.
We can also make use of the \citet{malygin14} gas opacities combined with the dust opacities
from \citet{semenov03} and compare these in Figure~\ref{Abb:kappaSchnitt}.

Note that the \citet{bl94} lacks water opacity lines just above
the dust destruction temperatures (M.\ Malygin, priv.\ comm.; see also Figure~1 of \citealp{ab06});
as a consequence, their opacities reach down to $\kapR\sim10^{-6}$~cm$^2$\,g$^{-1}$
for $\rho=10^{-11}$~g\,cm$^{-3}$, where $\kapR$ is the Rosseland mean,
some four orders of magnitude smaller
than in more recent calculations \citep{freed08,malygin14}.
Figure~\ref{Abb:ungefSchPopSynth_kalt} shows that for low masses and accretion rates,
the shock temperature should be $\TSch\lesssim1500$~K, in which case the dust is not destroyed 
and the opacity is relatively high.
When higher temperatures are reached, the opacity is lower by orders of magnitude,
so that the total (gas and dust) Rosseland opacities
range from $\kapR\sim10^{-2}$ to 10~cm$^2$\,g$^{-1}$ overall.
This provides approximate values when considering constant opacities.
Since 1-$T$ requires only the Rosseland mean, the subscript on $\kapR$
will be dropped hereafter.

\subsection{Quantities to be measured}

\subsubsection{Efficiencies}
 \label{Theil:etaDef}
The main goal of this study is to determine the radiative loss efficiency of the accretion shock.
There are several ways of defining this.
The classical definition, $\etaklassisch$,
indicates what fraction of the inward-\Ae{directed} kinetic energy flux
is converted into a jump in outgoing radiative flux \citep[e.g.,][]{hartmann97,baraffe12,zhu15}.
This kinetic-energy luminosity is \Ae{at most}
\begin{align}
 \Lkin &= 4\pi\RP^2 \frac{1}{2} \varrho v^3 = \frac{1}{2}\MPunkt v^2\\
      &\approx \frac{G\MP\MPunkt}{\RP},
\end{align}
where $\MPunkt=4\pi r^2 \varrho v$ is the mass accretion rate (neglecting the sign of $v$)
and the last expression is valid for free fall from a large radius. %
Therefore, the energy actually radiated away at the shock is written as
\begin{equation}
 \LAkk = \etaklassisch  \frac{G\MP\MPunkt}{\RP}
\end{equation}
and it is usually assumed that $\etaklassisch=100$~percent (i.e., full loss).
This is called `cold accretion'.
Note that this $\etaklassisch$ corresponds to the quantity
$(1-\eta)$ of \citet{spiegel12}, $\alpha_{\rm h}$ of \citet{morda12_I},
$(1-\alpha)$ of \citet{hartmann97}, and $X$ of \citet{commer11}.

We present here and use a second definition based on the total energy available.
This efficiency $\etaphys$
measures what fraction of the total energy flowing towards the planet
actually remains below the shock, i.e., is \Ae{absorbed} by the \Ae{embryo}:
\begin{equation}
\label{Gl:etaphys_DeltaLE}
 \etaphys \equiv \frac{ \EPkt(\rmax) - \EPkt(\rSchock^-) }{ \EPkt(\rmax) },
\end{equation}
where $\rSchock^-$ is immediately downstream of the shock
and \Ae{the outer edge of the computation domain} $\rmax$
\Ae{is used as a proxy for} the accretion radius $\RAkk$ corresponding
to the location of the nebula.
The material-energy flow rate is defined as
\begin{equation}
\label{Gl:EPkt(r)}
 \EPkt(r) \equiv -|\MPunkt|\left[\ekin(r) + h(r) + \Delta\Phi(r,\rSchock)\right],  %
\end{equation}
where $\ekin=\frac{1}{2}v^2$, $\eint$, and $h=\eint+P/\rho$ are respectively the
kinetic \Ae{energy},
internal energy density, and the enthalpy \Ae{per unit mass} and $\Phi$ the external potential.
The $\Delta\Phi$ term in Equation~(\ref{Gl:EPkt(r)}) accounts for the work done by the potential
on the gas down to the shock, with \Ae{the potential difference from $r_0$ to $r$ given by}
\begin{equation}
 \label{Gl:DeltaPhi}
 \Delta\Phi(r,r_0) = -G\MP\left(\frac{1}{r} - \frac{1}{r_0}\right).
\end{equation}
Thus $\etaphys$ measures how much of the incoming energy flow $\EPkt(\rmax)$ in the gas
is still flowing inward \Ae{once it has passed through} the shock;
if both are equal ($\EPkt(\rSchock^-)=\EPkt(\rmax)$),
$\etaphys=0$ and the accretion would be thought of as `hot'.
If in the other extreme case none of the energy traverses the shock, $\etaphys=100$~percent,
implying that the energy must have been entirely converted to outward-traveling radiation.
\Ae{This therefore automatically reflects the fact that the (non-)heating
of the planet is determined by the \textit{imbalance} between the amount of kinetic energy
converted to internal energy and the re-emitted radiation.}
By energy conservation, the numerator \Ae{of $\etaphys$} should be equal
to the difference $\EPkt(\rSchock^+)-\EPkt(\rSchock^-)$
between the material energy \Ae{flow rate directly} across the shock.
\Ae{This is true} for a zero-temperature gas (infinite Mach number),
\Ae{for which} the potential energy is entirely converted
in kinetic energy by the external potential.
For finite temperatures, however, a (small) pressure gradient builds up ahead of the shock;
in this case, only part of the change in potential energy serves to increase the kinetic energy,
the remainder going into internal energy and thus, outside of phase transitions, into pressure.

Also by energy conservation, $\Delta\EPkt(\rSchock^\pm)$ measured in the shock frame should be equal (up to a sign)
to the change in the luminosity $\Delta L$ across the shock.
However, in the case that the \Ae{radiative} precursor \citep{zeldovich67} is contained
within the accretion region---roughly the Hill sphere---,
it is not true anymore that $\EPkt(\rSchock^+)=\EPkt(\rmax)$.  %
In fact, the luminosity upstream of the precursor \Ae{can be} smaller than downstream  %
(i.e., the planet is invisible, at least in the grey approximation), which would lead to a negative efficiency
if using $\Delta L$.
Thus, $\Delta\EPkt$ is a more useful numerator because it is intuitive and applicable both when
the precursor reaches to $\rmax$ and not.

Note finally that the definition of Equation~(\ref{Gl:etaphys_DeltaLE}) 
\Ae{
takes into account the fact that
even if
the entire kinetic energy is converted to luminosity,
the net efficiency can still be zero} if this radiation is absorbed by the incoming material.
This was seen by \citet{vaytet13} in the case of Larson's second core
and estimated by \citet{baraffe12} to be the case at high accretion rates in the context
of magnetospheric accretion onto stars.

Thus, we will focus in this study on the efficiencies as defined above:
on the classical, `kinetic' efficiency $\etaklassisch$, which makes a direct statement
about the energy conversion at the shock, with $\etaklassisch<100$~percent
for either an isothermal shock at Mach number $\Mach\lesssim2.5$ \citep{commer11}
or an non-isothermal shock;
and on the `physical' efficiency $\etaphys$, which indicates how much the upstream gas
is able to recycle the energy liberated at the shock \citep{drake06}.

\subsubsection{Post-shock entropy}
The post-shock temperature and thus entropy depend on the thermal profile of the layers below the shock,
which are expected to adjust to carry the luminosity from deeper down \citep{paxton13}. %
Since however we do not attempt to predict this luminosity accurately with our set-up of a truncated atmosphere,
the reported temperature values will serve only as an indication.
Moreover,
there is a non-trivial relationship between the post-shock entropy values
and their influence on the entropy of the planet's deep adiabat;
in particular, the post-shock material does not simply set, weigthed by mass,
the interior entropy.
This question is the subject of separate studies (\citealp{berardo16}, Marleau et al., in prep.),
which however require the obtained post-shock entropies as boundary conditions.

\section{Results: radial profiles and efficiencies}

\begin{figure*}
 \epsscale{1.2}
 \plotone{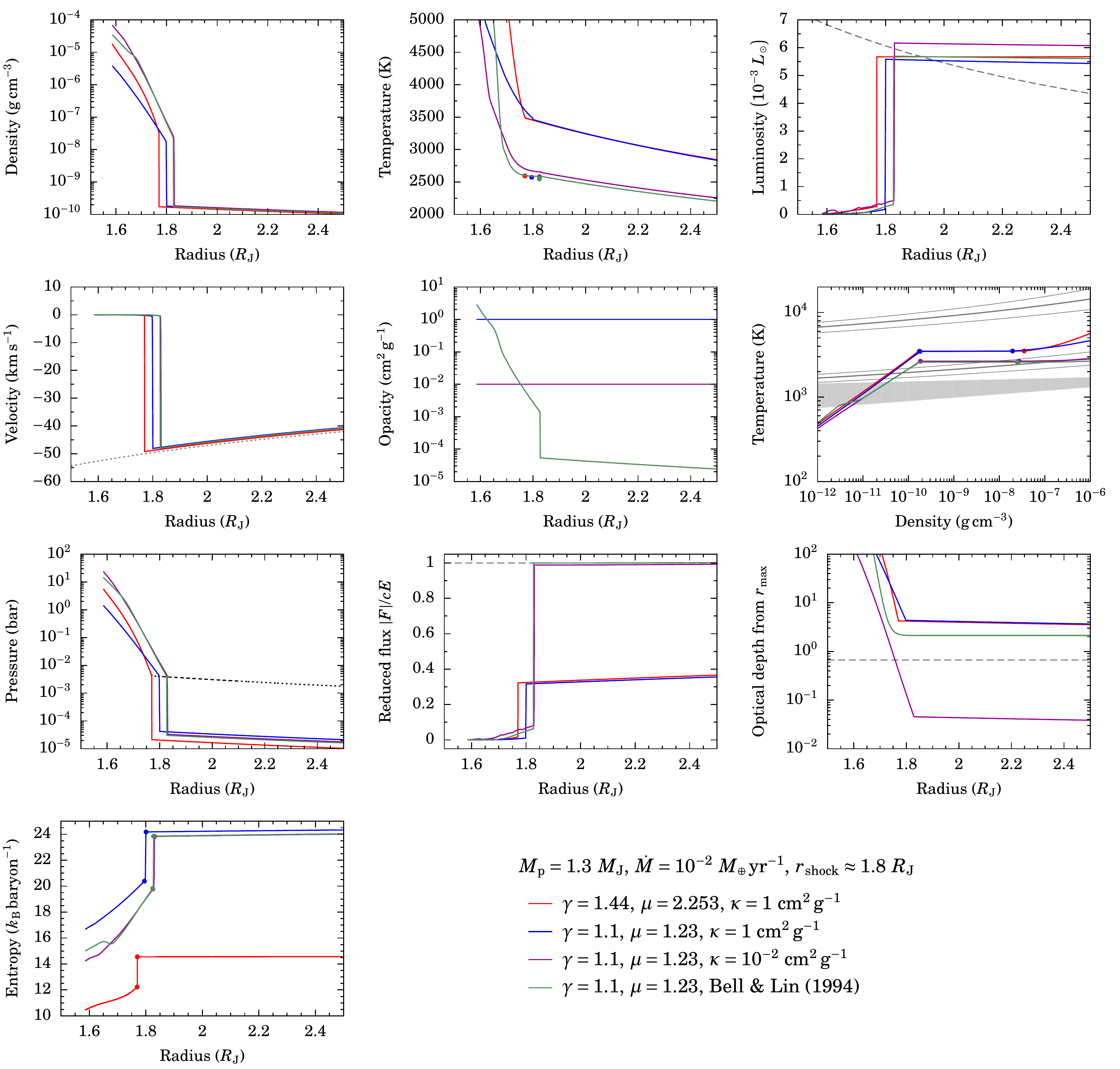}
 \caption{
 Detailed shock profiles for simulations with $\MP=1.3~\MJ$, $\rSchock\approx1.8~\RJ$, and $\MPunkt=10^{-2}~\ME\,\rm{yr}^{-1}$
 using a constant equation of state and constant opacities or the \citet{bl94} opacities (\textit{see legend}).
 The simulation grids extend to 0.7 of the accretion radius $\RAkk\approx250~\RJ$ but only the inner region is shown.
The axis labels describe the quantities shown, and only a few comments are needed:
The \textit{temperature} panel also shows %
the lower bound estimate of Equation~(\ref{Gl:TisothkonstLetakin100}; \textit{filled dots});
in the \textit{luminosity} panel, the maximal accretion luminosity $\LAkkmax(r)=\frac{1}{2}\MPunkt v(r)^2\approx G\MP\MPunkt/r$ is shown at
every radius (\textit{gray dashed curve});
the \textit{velocity} panel also displays %
the free-fall velocity from $\RAkk$ (the same for all simulations; \textit{grey dotted line});
in the \textit{opacity} panel, simulations with constant opacity overlap;
in the \textit{temperature--density phase diagram}, the solid dots mark the up- and downstream conditions of shock, %
the solid lines show contours of 10, 50, and 90~percent atomic hydrogen (relative to the hydrogen species),
and the grey region hightlights where the dust is being destroyed, with $\kappa\sim1$~cm$^2$\,g$^{-1}$ below
and $\kappa\sim10^{-6}$--10$^{-3}$~cm$^2$\,g$^{-1}$ above;
the \textit{pressure} panel also displays the ram pressure $\Pram=\rho v^2$, the same for all simulations (\textit{dashed curve}); %
and the \textit{entropy} is computed self-consistently from Equation~(\ref{Gl:EntroSackurTetrode}).
 }
\label{Abb:Vergleich-1.259MJ_MPkt1e-2_rSchockca.1.7RJ_Artikel1}
\end{figure*}

We have performed a large number of simulations, varying physical parameters (mass, radius, accretion rate)
but also computational or numerical settings (technique for accreting gas into the domain,
outer temperature boundary condition, resolution, Courant number, etc.).
For the latter, we select the most stable set-up (as described in Section~\ref{Theil:Kode} above),
and present results for a typical combination relevant to core accretion formation calculations \citep{boden00,morda12_I}.
We look at the properties of the accretion shock for $\MP=1.3~\MJ$, $\MPunkt=10^{-2}~\ME\,\rm{yr}^{-1}$, and $\rSchock\approx1.8~\RJ$.
The Bondi, Hill, and resulting accretion radius according to Equation~(\ref{Gl:RAkk})
are $\RBondi\approx4200~\RJ$, $\RHill\approx800~\RJ$, and $\RAkk\approx250~\RJ$
(for $\TNeb=150$~K and a solar-mass star, which however does not affect $\RAkk$ strongly).
Figure~\ref{Abb:Vergleich-1.259MJ_MPkt1e-2_rSchockca.1.7RJ_Artikel1} shows the detailed structure of the accretion flow
near the shock for $\kappa=10^{-2}$ and 1~cm$^2$\,g$^{-1}$, as well as with the \citet{bl94} opacities.
For the EOS, we consider a hydrogen--helium mixture with a helium mass fraction $Y=0.25$
and cases where hydrogen is everywhere molecular ($\mu=2.353$, $\gamma=1.44$) or atomic ($\mu=1.23$, $\gamma=1.1$).
The radial structures are as expected and show a number of typical features, which we discuss in the following.

 \subsection{Density, velocity, and pressure}
 \label{Theil:rho v P}
 The density and velocity reveal gas almost exactly free-falling onto %
 a nearly hydrostatic atmosphere abruptly cut off at the shock.
 The mass in the total domain, dominated by the post-shock region,
 is typically $\Delta M\sim10^{-4}~\ME$, making perfectly justified the neglect of the self-gravity
 of the gas.
The density jumps at the shock by a factor $\rho_2/\rho_1\sim200$,
where $\rho_2$ and $\rho_1$ are the post- and pre-shock density.
\Ae{Thanks to the transport of energy by radiation,}
this is a much larger compression than \Ae{the infinite-Mach number limit} for a hydrodynamical shock,
where $\rho_2/\rho_1=(\gamma+1)/(\gamma-1)\approx4$ to 20 \Ae{for $\gamma=\frac{5}{3}$ to~1.1}
\citep[e.g.,][]{mihalas84,commer11}.
As it falls deeper in the potential well of the planet, the gas slows down to slightly sub-free-fall speeds
due to the pressure gradient caused by the increasing temperature and density.

The post-shock pressure is given very accurately by the ram pressure of the incoming gas,
\begin{equation}
 P_{\rm post} = \Pram = \varrho v^2.
\end{equation}
This differs slightly from the strong-shock (high-Mach-number), non-radiating case
where $P_{\rm post} = 2/(\gamma+1) \varrho v^2$ \citep[][his Equation~4.18]{drake06},
as we verified with a simulation using \Ae{a higher} $\gamma=5/3$ to increase the difference.

\K{
}

\subsection{Optical depth, reduced flux, and radiation regime}
 \label{Theil:Deltatau,fred}
We begin by discussing the reduced flux. %
The reduced flux
\begin{equation}
 \label{Gl:fredDef}
 \fred\equiv \Frad/(c\Erad)
\end{equation}
is a \textit{local} measure of the extent
to which radiation is streaming freely ($\fred\rightarrow1$)
or diffusing ($\fred\rightarrow0$).
This is thus the more correct, physical measure of what is often loosely termed the optical depth,
as discussed below.
The reduced flux, radiation quantity $R$, and flux limiter $\lambda$
are related in general by $\fred=\lambda(R)R$.
Note that the effective speed of propagation of the photons is $\ceff=\fred c$.

Next, we consider the optical depth.
For
a free-fall profile with $\RAkk\gg\rSchock$ (so that $\rho\propto r^{-3/2}$) and a radially sufficiently constant opacity
($\kappa\propto r^{\alpha}$ with $|\alpha|\ll \frac{1}{2}$), the optical depth to the shock is
\begin{subequations}
\begin{align}
  \Delta\tau &=\int_\rSchock^\infty\kappa(r)\rho(r) {\rm d}r\\
             &=2\kappa\rho\rSchock~~~\textnormal{(const.~$\kappa$)}, %
\end{align}
\end{subequations}
where $\kappa\rho$ is evaluated at the shock\footnote{This
	justifies (within a factor of a few) the estimate
	$\kappa\varrho\rSchock$ of \citet{stahlerI} for the optical depth \textit{upstream}
	(and not downstream, as the formula might at first suggest) of the accretion shock
	in the context of Larson's second core.
	The estimate is certainly rough for a non-constant opacity
	but at least it does estimate the optical depth in the correct (upstream) direction.
A similar expression is used by \citet{morda12_I} in their boundary conditions.
}.
For the data of Figure~\ref{Abb:Vergleich-1.259MJ_MPkt1e-2_rSchockca.1.7RJ_Artikel1},
$\rho=1.5\times10^{-10}$~g\,cm$^{-3}$ upstream of the shock,
so that $\Delta\tau=3.9\left(\kappa/1~\mbox{cm$^2$\,g$^{-1}$}\right)$.
This agrees very well with the actual optical depths (measured from $\rmax=0.7\RAkk\approx250~\RJ$)
in the constant-opacity cases.
For the simulation with the \citet{bl94} opacities,
the estimate is moderately accurate,
if one takes for $\kappa$ not the actual pre-shock value ($\kappa\sim10^{-5}$~cm$^2$\,g$^{-1}$, set by the gas)
but rather a typical value ($\kappa\sim1$~cm$^2$\,g$^{-1}$) in the outer regions ($r\gtrsim40~\RJ$),
where the dust is not destroyed.
Also for non-constant opacities, then, the optical depth to the shock will be roughly given by %
\begin{align}
 \Delta\tau \sim 3 \left(\frac{\kappa}{1~\mbox{cm$^2$\,g$^{-1}$}}\right) & \left(\frac{\MPunkt}{10^{-2}~\ME\,{\rm yr}^{-1}}\right)\notag\\
                      &\times\sqrt{ \left(\frac{1~\MJ}{\MP}\right) \left(\frac{2~\RJ}{\rSchock}\right) },\label{Gl:DeltaUngef}
\end{align}
using that $\MPunkt=4\pi r^2\rho v$.
Since the nebula should always be at temperatures lower than the dust destruction temperature $\TZerst\approx1500$~K,
the high opacity of the dust will always contribute to the optical depth.
Therefore, independently of whether dust is destroyed in the inner parts of the flow, close to the shock,  %
the opacity to insert in Equation~(\ref{Gl:DeltaUngef}) should be of order $\kappa\sim1$~cm$^2$\,g$^{-1}$.

Secondly,
\Ae{writing $\Erad=\Frad/(c\fred)\propto L/(r^2\fred)$, it is clear that}
when $L/\fred$ is locally spatially constant ($L/\fred\propto r^{\beta}$ with $|\beta|\ll 2$),
the radiation quantity $R=1/(\kappa\rho)\left|\partial\ln\Erad/\partial r\right|$ %
is given by $R\approx 2/(\kappa\rho r)$ where $\kappa\rho$ and $r$ are evaluated locally\footnote{It  %
	may seem surprising that the local quantity $R$ depends on an absolute coordinate $r$
	but this is in fact a simple consequence of the (spherical) geometry.
	}.
This result applies in general, independently of the radiation regime (diffusion or free streaming).

Combining 
these observations leads to the result that, in the case of constant opacity and $L/\fred$,
the reduced flux upstream of the shock is %
\begin{subequations}
\label{Gl:fredvonDeltau_approx}
\begin{align}
 \fred(\rSchock^+) &= \lambda\left(\frac{4}{\Delta\tau}\right)\times\frac{4}{\Delta\tau}\label{Gl:fredvonDeltau}\\
       &\approx \left(\frac{3}{4}\Delta\tau+1\right)^{-1},
\end{align}
\end{subequations}
where the second line would be an equality for the simple flux limiter $\lambda=1/(3+R)$ \citep{leblancwilson70,lever84,ensman94}.
We will return to this result in Section~\ref{Theil:DiskussionT}.
For a non-constant opacity, Equation~(\ref{Gl:fredvonDeltau_approx}) provides
in fact an approximate lower bound of $\fred(\rSchock^+)$ given $\Delta\tau$ or vice versa:
indeed, a low opacity in front of the shock will drive down $\fred$
(compared to the prediction of Equation~(\ref{Gl:fredvonDeltau_approx}))
but without decreasing much the total optical depth.
This highlights the conceptual independence between the (non-local) optical depth
and the (local) radiation transport regime (free-streaming or diffusion).

The optical depths from the shock out to $\rmax\approx\RAkk$
are $\Delta\tau\approx2$--5 for all except the $\kappa=10^{-2}$~cm$^2$\,g$^{-1}$ simulation,
which has $\Delta\tau\approx3\times10^{-2}$.
(A comparison run with the dust opacities of \citet{semenov03} and the gas opacities of \citet{malygin14}
yielded very similar profiles and optical depths.)
\Ae{In the $\kappa\neq10^{-2}$~cm$^2$\,g$^{-1}$ simulations,}
the shock would therefore be called `optically thick'.
However, the effective speed of light $\ceff=\fred c\gtrsim0.3c$ throughout the flow (see below),   %
which is still orders of magnitude larger than the gas flow speed $v\sim10^{-4}c$.
This is the regime \citet{mihalas84} term `static diffusion'.
Therefore, the radiation is able to diffuse into the incoming gas, heating it up out to the edge of the computation grid,
near the accretion radius. %
In other words, the shock precursor is larger than the Hill radius,
which implies that the radiation should be able to escape from the system
to at least the local disk.
In this sense, the shock for these parameter values is 
an optically thick--thin shock (down- and upstream, respectively)
in the classification of \citet{drake06}.
\Ae{That despite the somewhat high optical depth the shock
is not equivalent to a hydrodynamical shock is already hinted at
by the large compression ratio
pointed out in Section~\ref{Theil:rho v P}. %
}

\K{    %
}

\subsection{Temperature}
 \label{Theil:DiskussionT}

\subsubsection{Shock temperature}

For all choices of $\kappa$ and the EOS ($\gamma$, $\mu$),
the temperatures immediately up- and downstream of the shock are essentially equal,  %
i.e., there is no jump in the temperature.
This is thus a \textit{supercritical} shock \citep{zeldovich67}, in which the downstream gas
is able to pre-heat the incoming gas up to the post-shock temperature.
Note that the 1-$T$ approach to the radiation transport used here cannot reveal the Zel'dovich spike
expected in the gas temperature.
This feature of radiation-\hspace{0pt}hydrodynamical shocks consists of a sharp increase of the gas temperature immediately
behind the shock, followed by a quick decrease in a `radiative relaxion region',
while the radiation temperature remains essentially constant
(\citealp{zeldovich67,mihalas84,stahlerI}; see \citealt{drake07} and \citealt{vaytetgonz13} for a more detailed description).
However, this is not of concern since this spike is very thin both spatially
(physically, a few molecular mean free paths, broadened in simulations to a few grid cells;
e.g., \citealp{ensman94,vaytetgonz13}; Marleau et al.\ in prep.) and in optical depth,
and below the Zel'dovich spike, the matter and radiation equilibrate again.
Therefore, the Zel'dovich spike should affect 
neither the post-shock temperature or entropy nor the shock efficiency.
A possible disequilibrium in temperatures just upstream of the shock
will be explored in a forthcoming publication. %

We find shock temperatures of $\TSchock\approx2500$~K for the cases
with a low pre-shock opacity ($\kappa=10^{-2}$~cm$^2$\,g$^{-1}$
or with \citet{bl94}),
but $\TSchock\approx3500$~K for the other two cases, both with $\kappa=1$~cm$^2$\,g$^{-1}$.
These temperature values (and their relatively large difference of 1000~K)
can be understood from an analytical estimate, presented next.
Firstly, one can always write
\begin{equation}
 \label{Gl:F_rSchock^+-}
 F(\rSchock^+) = F(\rSchock^-) + \etaklassisch\frac{1}{2}\rho \vSch^3,
\end{equation}
where here $\rho$ is the density just ahead of the shock,  %
$\vSch$ is the velocity at the same location,  %
and $\etaklassisch$ is the `kinetic-energy loss efficiency', discussed in Section~\ref{Theil:eta}.   %
In general, the flux on either side of the flux is $F(\rSchock^\pm) = \fred^\pm c a T^4(\rSchock^\pm)$,
where $\fred^\pm\equiv\fred(\rSchock^\pm)$ and $a$ is the radiation constant,
related to the Stefan--Boltzmann constant $\sigSB$ by $ac=4\sigSB$.
Note that here $\fred=\Frad/c\Erad$ should be negative (one usually implicitly takes the norm)
if the downstream radiation is flowing inward ($\Frad<0$).
For an isothermal shock at $\TSchock$, Equation~(\ref{Gl:F_rSchock^+-}) then implies that
\begin{equation}
\label{Gl:TSchock_rauh}
 \sigSB \TSchock^4 = \frac{\etaklassisch}{4\Delta\fred} \frac{\rho \vSch^3}{2},
\end{equation}
where $\Delta\fred\equiv\fred^+-\fred^-$.
Combining with Equations~(\ref{Gl:DeltaUngef}) and~(\ref{Gl:fredvonDeltau_approx}) yields the estimates
for an isothermal shock
\begin{subequations}
\begin{align}
 \TSchock(\Delta\tau\ll1) \approx  &~2315~{\rm K} \left(\frac{\rSchock}{2~\RJ}\right)^{-3/4} \notag \\  %
                                   &\left(\frac{\MPunkt}{10^{-2}~\ME\,{\rm yr}^{-1}}\right)^{1/4} 
                                     \left(\frac{M}{1~\MJ}\right)^{1/4}\\
 \TSchock(\Delta\tau\gg1) \approx  &~2710~{\rm K} \left(\frac{\rSchock}{2~\RJ}\right)^{7/8} \left(\frac{\kappa}{1~{\rm cm}^2\,{\rm g}^{-1}}\right)^{1/4} \notag \\  %
                                   &\left(\frac{\MPunkt}{10^{-2}~\ME\,{\rm yr}^{-1}}\right)^{1/2} 
                                     \left(\frac{M}{1~\MJ}\right)^{1/8} \label{Gl:TisothkonstLetakin100},
\end{align}
\end{subequations}
where a $\left({\etaklassisch}\right)^{1/4}$ factor was left out on the right-hand sides
since we find it is $\approx1$ (see Section~\ref{Theil:eta}).
The first expression used that, by Equation~(\ref{Gl:fredvonDeltau_approx}), $\Delta\tau\ll1$ implies $\fred^+\approx 1$,
and further took $\fred^-\ll\fred^+$.
The second case is somewhat crude for non-constant opacities.
This assumes a constant luminosity in the shock's near upstream vicinity.
Since the post-shock region is very dense, $\fred^-$ is small;
this is equivalent to neglecting the downstream luminosity,
which is related in a non-trivial way to the interior luminosity of the planet
(\citealp{berardo16}; Marleau et al., in prep.).

The filled circles in Figure~\ref{Abb:Vergleich-1.259MJ_MPkt1e-2_rSchockca.1.7RJ_Artikel1}
show the lower bound of Equation~(\ref{Gl:TisothkonstLetakin100}).
The simulations with a low pre-shock opacity \Ae{($\kappa=10^{-2}$~cm$^2$\,g$^{-1}$} or with \citet{bl94})
have $\fred\approx1$ upstream of the shock and
indeed have a temperature given by Equation~(\ref{Gl:TisothkonstLetakin100}),
whereas in the other cases a higher temperature is needed to carry a similar luminosity.
The difference is quite large and nearly 1000~K.
One way of thinking about this is that the effective speed of light is lower than $c$,
so that $\Erad$ must increase in order to reach the same $\Frad=c_{\rm eff}\Erad$.

Interestingly, the molecular- and atomic-hydrogen cases lead to a very similar temperature
$\TSchock=3500$~K. %
The phase diagram indicates that the atomic-hydrogen simulation with $\kappa=1$~cm$^2$\,g$^{-1}$
is self-consistent,
but that the case with atomic hydrogen and detailed opacities leads to temperatures
and densities where the dissociation process (and thus a varying $\mu$ and $\gamma$)
would be important.
One can already anticipate the result that, for an isothermal shock,
the hydrogen should \textit{recombine} in part through the shock (Marleau et al., in prep.)
since at fixed temperature the abundance of H$_2$ increases with density.
Note that
\citet[][their Equation~24]{stahlerI} present an estimate similar to Equation~(\ref{Gl:TSchock_rauh})
in the context of stellar accretion.
Their assumptions about the reprocessing of shock photons\footnote{They
	assume that half of the photons generated at the shock move inward,
	and the other half outward; in turn, one half of this outward-moving radiation
	is assumed to be reradiated inward by an absorbing layer ahead of the shock.
	If one ignores the contribution from the interior luminosity,
	this implies that $\etaklassisch=25$~percent.
	However, it seems to us that one needs radiative transfer calculations
	such as the ones presented here (or using more detailed radiation transport
	as in \citealp{drake07}) to justify this accounting. %
	}
imply that,
when $\Frad(\rSchock^-)\ll\Frad(\rSchock^+)$ and neglecting their $T_d$ term, 
$\Delta\fred\approx\fred^+\approx1/3$ automatically.
\citet[][their Equation~22 or~53]{commer11} give a formula
\Ae{similar to Equation~(\ref{Gl:TSchock_rauh}) in the limiting case $\etaklassisch=1$}
but %
do not include the \Ae{factor} $1/(4\Delta\fred)$.
\Ae{This is because they equate the temperature at the shock
with the \textit{effective} temperature needed to radiate away the kinetic energy,
increasing the temperature estimate by $\approx40$~\%\ (a factor $4^{1/4}\approx1.4$),
or $\approx1200$~K for $\TSch\approx3000$~K}.

\subsubsection{Temperature profile}
Equation~(\ref{Gl:fredDef}) implies that,
if the luminosity and the reduced flux are radially roughly constant,
$T\propto r^{-1/2}$ since $L=4\pi r^2 \Frad$,
independently of the optical depth to the shock.
This is the case for the constant-$\kappa$ simulations
but not so for the tabulated opacities (at larger radial distances than shown).

Note that if the temperature increased solely due to adiabatic compression,
i.e., at constant entropy in the absence of radiation transport,
we would have $T\propto \varrho^{\gamma-1}\propto r^{-1.5(\gamma-1)}$,
i.e., $T\propto r^{-0.15}$ or $T\propto r^{-0.66}$ for $\gamma=1.1$ or $1.44$, respectively.
Thus, when $T\propto r^{-1/2}$, entropy decreases inward if $\gamma>4/3\approx1.33$. %

\subsection{Entropy}
To compute the entropy, we use the Sackur--Tetrode equation \citep[e.g.,][and references therein]{mc14,berardo16}
for an ideal gas composed of H$_2$ and He or H and He:
\begin{subequations}
 \label{Gl:EntroSackurTetrode}
\begin{align}
 S_{\textrm{H}_2\mbox{--}\textrm{He}} =&   %
     \,8.80 + 3.38\log_{10}\left(\frac{T}{1000~{\rm K}}\right)\notag \\
                                                 &- 1.01\log_{10}\left({P\over 1\ {\rm bar}}\right),\\
 S_{\textrm{H}\mbox{--}\textrm{He}} =&
     \,13.47 + 4.68\log_{10} \left(\frac{T}{1000~{\rm K}}\right)\notag \\
                                                 &- 1.87\log_{10}\left({P\over 1\ {\rm bar}}\right),
\end{align}
\end{subequations}
respectively, using $Y=0.243$, \Ae{and where the entropies are in units
of Boltzmann's constant per baryon, $\kB/\mbox{baryon}$}.
In Figure~\ref{Abb:Vergleich-1.259MJ_MPkt1e-2_rSchockca.1.7RJ_Artikel1}, we see that the entropy \textit{decreases}
across the shock by $|\Delta S|\approx2.5$ and 4.0~$\kB/\mbox{baryon}$  %
for the molecular and atomic case, respectively.
\Ae{(In general but for constant $\gamma$ and $\mu$,
the jump in entropy at an isothermal shock is $\Delta S = -2.303/\mu\times \log_{10}(\gamma\Mach^2)$
in units of $\kB/\mbox{baryon}$.)}
That the entropy decreases through this shock is actually in agreement
with the statement that entropy increases across a hydrodynamical shock.
Indeed, once it arrives at the radiative shock found here, the gas has already seen its entropy increase
from the value far outside of the precursor.
(In the case that the precursor is larger than the simulation domain,
as applies for these simulations, this `far-field' value cannot be obtained directly. %
\Ae{However, already at $\rmax$ is the entropy much lower than downstream of the shock.})
Thus the radiative shock which is the subject of this work can be thought as being embedded in a usual hydrodynamical shock,
a `shock within a shock' \citep{mihalas84},
or a hydrodynamical shock as being a radiative shock with an infinitely \Ae{or unresolved} thin precursor.
Separate test simulations with extremely high opacity values ($\kappa=10^2$~cm$^2$\,g$^{-1}$),
such that the precursor is contained in the simulation domain,
confirm that the post-shock entropy is higher than the entropy far away from the shock.

The post-shock entropies are respectively $S\approx12$ and 20~$\kB/\mbox{baryon}$ for the molecular and atomic cases.
Compared to the range of entropies seen for cold starts to hot starts ($S\approx8$--14~$\kB/\mbox{baryon}$; \citealp{marl07,spiegel12,morda13}),
this is an extremely large difference, which is due mostly to the different mean molecular weights.
Moreover, it highlights the importance of using a self-consistent EOS which follows in particular
the dissociation of hydrogen.
However, the entropy values do not depend sensitively on the precise opacity
(see Figure~\ref{Abb:Vergleich-1.259MJ_MPkt1e-2_rSchockca.1.7RJ_Artikel1}).

Finally, it is important to remember that these entropy values are meant to be rather indicative at this stage.
First of all, they are not entirely self-consistent with the probable state of the hydrogen
in all parts of the domain (see the phase diagram in Figure~\ref{Abb:Vergleich-1.259MJ_MPkt1e-2_rSchockca.1.7RJ_Artikel1}).
Second of all, what they actually imply for the post-formation entropy
needs to be worked out separately, with a study of the post-shock settling region
and its coupling to the planet interior (\citealp{berardo16}, Marleau et al., in prep.).

\subsection{Luminosity}
The luminosity increases from the imposed $L=0$ value at $\rmin$ to the shock
where it jumps by a finite amount $\Delta L$, then decreasing with radius.
The value of $L$ downstream of the shock reflects in part the cooling of the layers below it,
and is set in reality also by (inefficient) convective energy transport, %
which we do not attempt to include in these simulations.
Thus the post-shock gas will probably have a different thermal history than
if the layers were allowed to sink further down into the planet
instead of stopping at most at $\rmin$.
Nevertheless, the obtained immediate post-shock luminosities are roughly $\Llinks\approx3\times10^{-4}~\LSonne$
and thus have values comparable to the (rough) internal luminosities of accreting planets (Mordasini et al., submitted).
Therefore, the inclusion of convection or similar changes to the temperature structure
should not lead to very different values for the post-shock region.

A general feature of these shock simulations is that $L$ decreases radially outward.
This is not due to absorption of the light with optical depth according to $L\propto \exp{(-\Delta\tau)}$,
as one might naively expect, but rather reflects energy conservation.
To derive this, we start with the total energy equation \citep[e.g.,][]{kuiper10},
\begin{equation}
\label{Gl:EnGlallg}
 \frac{d\Etot}{dt} + \nabla\cdot([\Ekin+H] v + \Frad) = \varrho v\cdot g,
\end{equation}
where the total energy volume density is $\Etot=\Ekin+H$,
with $\Ekin=\frac{1}{2}\rho v^2$,
and the enthalpy is $H=\Eint+P$ for an internal energy density $\Eint$.
For a constant EOS, $\Eint =\rho\cV T=\rho/(\gamma-1)\times\kB T/(\mu\mH)=1/(\gamma-1)\times P$,
where $\cV$ is the heat capacity.
It is easy to verify that the thermal timescales are much shorter
than the dynamical timescales, so that the flow is in steady state
and the time derivative $d\Etot/dt$ can be neglected.
Also, $\MPunkt$ is constant radially.
Remembering that $\nabla\cdot F = 1/r^2d/dr(r^2 F)$ for a vector $F$,
Equation~(\ref{Gl:EnGlallg}) becomes
\begin{equation}
 \label{Gl:dLdr}
 \frac{dL}{dr} = \MPunkt \frac{dh}{dr} + \MPunkt \frac{d}{dr}
 \left(\frac{1}{2}v^2 - \frac{G\MP}{r}\right),
\end{equation}
where $h=H/\rho$ is the specific enthalpy per mass and is $h=\gamma/(\gamma-1)\kB T/(\mu\mH)$ for a constant EOS.
The accretion rate $\MPunkt$ was taken to be positive here, i.e., $\MPunkt=|4\pi r^2\rho v|$,
\Ae{and one can trivially replace $G\MP/r$ by $G\MP\left(1/r-1/\RAkk\right)$.}
\Ae{If the second term on the righthand side of Equation~(\ref{Gl:dLdr}) is small,}    %
Equation~(\ref{Gl:dLdr}) shows that the radial decrease in $L$
is \Ae{mostly} due to the inward increase in enthalpy.
Therefore, it is not an explicit function of the optical depth,
although $T(r)$ and thus $h(T)$ are indirectly set by the opacity.
Note that this derivation is valid for a general EOS (with variable effective $\gamma$)
and also does not depend on the opacity being constant.

Equation~(\ref{Gl:dLdr}) can be integrated to yield,
\Ae{when the second term on the righthand side of Equation~(\ref{Gl:dLdr}) is negligible},
\begin{equation}
 \label{Gl:L-Abfall(r)}
 L(r)-\Llinks = \Delta L(\rSchock)\left[1 - \frac{\MPunkt\Delta h(r)}{\Delta L(\rSchock)}\right],
\end{equation}
where %
$\Delta L(\rSchock) = \etaklassisch\LAkkmax$ is the jump in luminosity at the shock,
and $\Delta h(r)\equiv h(\rSchock)-h(r)$
is the change in enthalpy relative to the shock,
with $\Delta h>0$ for outwards decreasing \Ae{enthalpy}.
This result seems plausible:
the inward enthalpy flux is comparable to the outward radiation flux
only when
the \Ae{infalling} gas absorbs a significant fraction of the radiation
\Ae{and thus decreases $L$}.
The maximal drop in luminosity \Ae{occurs for $h(\rmax)\ll h(\rSchock)$,
i.e., when the effective nebula temperature $\TNeb\ll\TSchock$.
This leads to $L(\rmax)-L(\rSchock^+)=-\MPunkt h(\TSchock)$.}  %

\subsection{Efficiencies}
 \label{Theil:eta}
\begin{figure*}
 \epsscale{1.2}
 \plotone{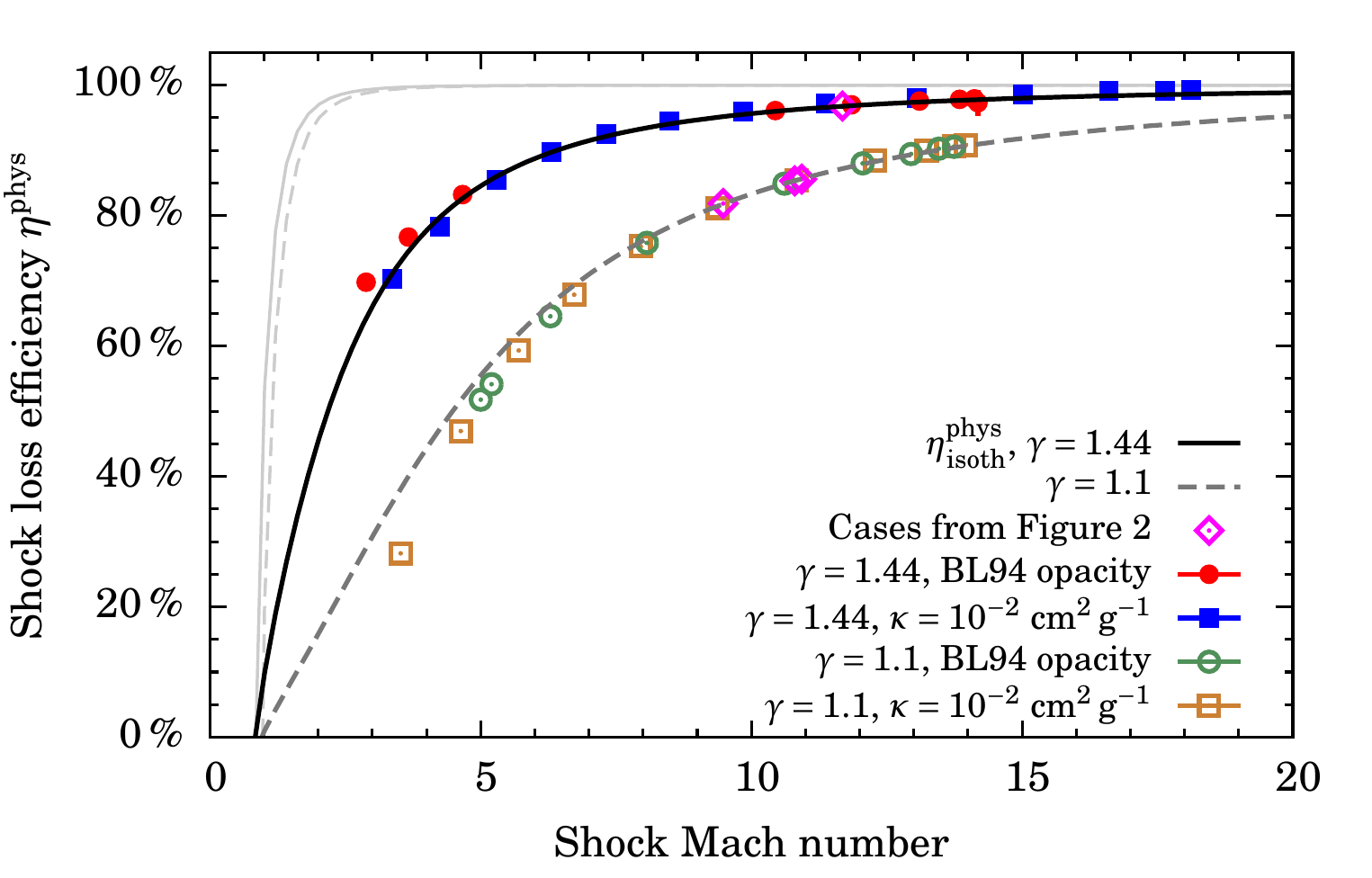}
 \caption{
 Physical loss efficiency $\etaphys$ of the radiative accretion shock
 (see Equation~(\ref{Gl:etaphys_DeltaLE})). %
 The limit $\etaphys=0$ corresponds to all the incoming energy being absorbed (no loss),
 while $\etaphys=100$~percent means that the kinetic energy of the gas entirely leaves the accretion flow
 onto the planet; see Section~\ref{Theil:etaDef}.
 The diamonds display the efficiency for the cases shown in Figure~\ref{Abb:Vergleich-1.259MJ_MPkt1e-2_rSchockca.1.7RJ_Artikel1}.
 The other points come from considering a range of accretion rates $\MPunkt=10^{-5}$--$10^{-2}~\ME\,{\rm yr}^{-1}$,
 masses $\MP\approx0.3$--10~$\MJ$, and shock locations $\rSchock\approx1$--$20~\RJ$.
Both constant and tabulated opacities are used as indicated in the legend.
The last four groups of points (see legend) all take $\mu=2.353$.
The results match the analytical result for an isothermal shock %
at the measured Mach number
(Equation~(\ref{Gl:etaisoth_klass}, \Ae{from \citealp{commer11}}), for $\gamma=1.44$ and $\gamma=1.1$
(\textit{solid black and dashed dark gray curves, respectively}).
Theoretical curves for the `kinetic efficiency' $\etaklassisch_{\textrm{isoth}}$
for an isothermal shock are shown for comparison (\textit{pale grey curves}).
 }
\label{Abb:eta}
\end{figure*}

Next we show in Figure~\ref{Abb:eta} the main result for the examples of Figure~\ref{Abb:Vergleich-1.259MJ_MPkt1e-2_rSchockca.1.7RJ_Artikel1},
the loss efficiency $\etaphys$ of the accretion shock.
We recall that $\etaphys=0$ would correspond to all the kinetic energy 
of the gas being absorbed by the planet and the gas \Ae{being accreted},
while $\etaphys=100$~percent would correspond to the entire kinetic energy
being radiated away when going out to the accretion radius $\RAkk$, roughly the Hill sphere
(here approximated by the outer radius near $\RAkk$).
Contrary to $\etaklassisch$, \Ae{$\etaphys$} takes into account
the energy recycling which occurs due to the incoming gas absorbing
the radiation liberated at the shock \Ae{(see Equation~(\ref{Gl:etaphys_DeltaLE})
and the discussion below Equation~(\ref{Gl:DeltaPhi}))}.
We find that they are $\etaphys\approx85$~percent for the atomic-hydrogen cases with different opacities,
and $\etaphys\approx95$~percent for the molecular case.
Thus, a fraction $1-\etaphys\approx5$--15~percent of the total incoming energy is added to the planet.
How significant this is for the energy budget of the planet can be assessed
by comparing $(1-\etaphys)\EPkt(\rmax)$ to the internal luminosity of the planet.
For these simulations, both are typically of the same order of magnitude,
implying that the accreting gas is able to heat the downstream region.
As mentioned above, how this then affects the entropy and luminosity
of the planet and their evolution will have to be studied separately.

We show also the efficiencies from simulations covering a range of accretion rates $\MPunkt=10^{-5}$--$10^{-2}~\ME\,{\rm yr}^{-1}$,
masses $\MP\approx0.3$--10~$\MJ$, and shock locations $\rSchock\approx1$--$20~\RJ$,
and varying again the opacity.
At the largest radii, efficiencies down to almost 20~percent are reached,
and to 99~percent at the other extreme.  %

By contrast, the `kinetic efficiency' is $\etaklassisch\approx100$~percent
(up to the numerical accuracy of the code given the resolution)
for all simulations shown in Figure~\ref{Abb:eta}, with $(1-\etaklassisch)\Etot$ smaller by orders of magnitude than $\LP$.
In other words, the entire kinetic energy is converted to an immediate jump in the luminosity
as the gas is brought to subsonic speeds through the shock.
However, a significant fraction does get reabsorbed in the accretion flow,
leading to the lower $\etaphys$ values.
Nevertheless, we find generally that the precursor is greater than the accretion radius,
which is of order of the Hill radius.
The optical depths from the shock to the Hill sphere are at most $\Delta\tau\sim30$,   %
and using a variable equation of state (which would yield other temperatures)
should not change this significantly.  %
We therefore expect the radiation to always be able to escape from the shock to the local disk (the nebula).

These numerical results can be compared to analytical theory for radiative shocks.
\citet[][his equation~7.82]{drake06} derived that the kinetic efficiency of a shock in which radiation pressure is negligible
is in general given by
\begin{equation}
 \etaklassisch \equiv \frac{\Delta F}{\frac{1}{2}\varrho_-{v_-}^3}= 1 +
                                     \frac{2}{(\gamma-1){\Mach}^2}\frac{\mathfrak{r}-1}{\mathfrak{r}}+\frac{\gamma+1}{\gamma-1}\frac{1}{\mathfrak{r}^2}.
\end{equation}
where $\mathfrak{r}\equiv \rho_2/\rho_1$ is the ratio of the post-shock to the pre-shock density.
For isothermal shocks (as we find here), $\mathfrak{r}=\gamma \Mach^2$,
and \citet{commer11} show that the efficiency is then %
\begin{equation}
\label{Gl:etaisoth_klass}
 \etaklassisch_{\textrm{isoth}}=1 - \frac{1}{\gamma^2\Mach^4}.
\end{equation}
\Ae{Thus a higher Mach number leads to a higher fraction of the incoming kinetic energy
being converted to radiation for an isothermal shock.
}
Since the total energy flux is
\begin{subequations}
\begin{align}
  \varrho v\etot &= \varrho v\left(\frac{1}{2}v^2 + h\right)\\
               &= \frac{1}{2}\varrho v^3\left(1 + \frac{2}{\gamma-1}\frac{1}{\Mach^2}\right), %
\end{align}
\end{subequations}
we can derive
that the physical efficiency, as measured by $\Delta L$ at the shock, is
\begin{equation}
\label{Gl:etaisoth_phys}
 \etaphys_{\textrm{isoth}} = \etaklassisch_{\textrm{isoth}} \times \left(1 + \frac{2}{\gamma-1}\frac{1}{\Mach^2}\right)^{-1}.
\end{equation}
Therefore, the physical efficiency is lower than the kinetic
since the former considers the heating of the radiative precursor.
In other words, not all radiation liberated at the shock can leave the planet,
and therefore gets incorporated in the planet's entropy.
Note that naively, one might expect in strongly supersonic flows ($\Mach=v/\cs\gg1$)
the internal energy (measured by $\cs^2$) to be negligible compared to the kinetic energy (measured by $v^2$),
but the $2/(\gamma-1)$ factor can make this assumption cruder than expected, especially for low $\gamma$ values;
for instance, when $\gamma=1.1$ and even with a high Mach number $\Mach=10$,
the factor $2/\Mach^2(\gamma-1)$ is $0.2$, i.e., a 20~per cent contribution. 

The Mach numbers range from $\Mach\approx3$--20, and
the $\etaphys_{\textrm{isoth}}$ curve is compared to the data in Figure~\ref{Abb:eta} for $\gamma=1.1$ and $\gamma=1.44$.
The agreement is excellent.
The deviation from the theoretical curve, seen for a few simulations,
is possibly due to small measurement errors related to the identification of the shock region,
and to inaccuracies in the measurement of the velocity at which the shock is spreading;
this speed becomes somewhat important (at the several-percent level) at low Mach numbers.
However, the overall agreement is excellent, independent of the opacity
and optical depth in the flow (not shown).

At least for the constant EOS used here, 
these simulations and other tests indicate that extreme parameter values
(e.g., $\MPunkt>10^{-1}~\ME\,{\rm yr}^{-1}$ or $\kappa>100$~cm$^2$\,g$^{-1}$)
would be needed to obtain a shock with a Mach number $\Mach\lesssim2$,
in which $\etaklassisch$ would clearly be lower than 100~percent.
Note that, while $\vFf\propto \sqrt{\MP}$, very small masses are not sufficient to obtain a lower Mach number
since $\Mach\propto v/\sqrt{T}$; at lower masses, the temperature in the pre-shock
region too is smaller, which does not let $\Mach$ get much lower than about 3.

\section{Discussion and Summary} %

We have studied spherically symmetric gas accretion onto a gas giant
during the detached runaway phase, when the gas falls freely
from the accretion radius (of order of the Hill radius) onto the planet, where it shocks.
We determine the radiative efficiency %
of the shock at the planet's surface and
argue that this should be defined with the total incoming energy flux,
i.e., taking both the kinetic but also the internal energy into account.
Even if, at a Mach number $\Mach=4$, an isothermal shock converts 100~percent
of the incoming kinetic energy into radiation,
\Ae{only 77~percent (40~percent) for $\gamma=1.44$ ($\gamma=1.1$) ultimately escape,
with 23~percent (60~percent) absorbed by the infalling gas
and therefore reaccreted to the system.}
This efficiency has 
direct observational consequences as it controls the amount of radiation
which leaves the planet and is possibly observable.  %
The efficiency is also important since the complementary fraction is carried through
the shock into the settling region, where the gas is being incorporated to the planet.
To the best of our knowledge,
the energetics of the shock have not yet been studied in detail as we have done,
yet are thought to be key in determining the post-formation thermal state of gas giants,
with several orders of magnitude of difference in the resulting luminosity
between the two extreme cases, hot and cold starts.

We have considered both constant and tabulated opacities \citep{bl94}
but have only used
a constant equation of state to concentrate on the shock physics.
Therefore, the numerical results are rather illustrative in a quantitative sense,
but the qualitative behavior of the radial profiles and the derived results
revealed a number of interesting features.
We find the following:
\begin{enumerate}
 \item The shock was observed always to be isothermal, which corresponds in the classical terminology
       to a supercritical shock \citep{stahlerI,mihalas84}.
 \item The effective speed of light of the escaping photons %
       is always much larger than the gas flow speed,
       so that the upstream region is in the `static diffusion' regime \citep{mihalas84}.
 \item Our radiation-hydrodynamics simulations confirm, over a large range of Mach numbers,
       the theoretical expression for the efficiency given by \citet[][our Equation~\ref{Gl:etaisoth_klass}]{commer11}.
 \item 
       Unrealistically high constant opacity values %
       were separately verified to be needed to cause the luminosity generated at the shock
       to be completely absorbed in the precursor, ahead of the shock region.
       For reasonable constant or tabulated opacities, all luminosity profiles
       are qualitatively similar, decreasing by some amount with increasing
       \Ae{distance and with a non-zero value at the outer edge (see last point below)}.
       An analytical formula is derived for the drop based on energy conservation
       and shows that, \Ae{roughly,} the decrease in luminosity is significant only if the incoming gas
       carries a significant amount of energy compared to the accretion luminosity.
 \item We generally find higher shock temperatures then predicted by 
       the usual estimate of the shock temperature, %
       Equation~(\ref{Gl:TisothkonstLetakin100}).
       We show analytically that this is a lower bound.
       The shock temperature being higher is due to the radiation
       of the pre-shock matter.
       (The difference between the actual and estimated temperature can be large---\Aeneu{near} 1000~K in our examples---,
       enough to possibly change the state of the gas significantly, from molecular to atomic.)
       This leads to lower Mach numbers and thus overall lower efficiencies of the shock.

 \item %
       The entropy was seen to decrease across the shock since it is in fact the \textit{radiative} shock
       embedded in the hydrodynamical shock; over the latter, the entropy does increase as expected.
       The decrease $\Delta S$ was found to be large, with $\Delta S\approx 1.5$--4~$\kB/\mbox{baryon}$
       for the examples considered.
       Thus the shock is very efficient in radiating away the entropy of the shocked gas.
       The post-shock values were seen to be clearly high ($S>12~\kB/\mbox{baryon}$),
       with the choice for the EOS making a significant difference.
       We however point out that the obtained densities and temperatures
       were not consistent with the assumed (constant) mean molecular weight
       and heat capacity.
       Therefore the entropy values, while consistent within the parameter choices for the simulations,
       should in general be expected to be different when using a non-constant complete equation of state.
       This will be the subject of Paper~II.
 \item For most of the formation parameter space,
       nearly all of the kinetic energy is radiated away at the shock, i.e., $\etaklassisch\approx100$~per cent.
       This is in agreement with the analytical formula of \citet{drake06} and \citet{commer11},
       which predicts $\etaklassisch\approx100$~percent for sufficiently high upstream Mach numbers ($\Mach\gtrsim3$).
       However, it is important to remember that the Mach number itself depends on the shock temperature,
       which is an \textit{outcome} of the simulations and can at best only be estimated beforehand.
 \item However, most importantly, we found that the physical (or ``planet-heating'') efficiency  %
       is usually smaller than 100~percent,
       with values down to $\etaphys\approx20$~percent for a reasonable range of parameter values.
       This energy flux coming into the planet is often comparable to or in fact much higher than
       its internal luminosity, suggesting that the accretion process can play an important role also energetically.
       The complementary fraction of the accretion luminosity should reach at least the Hill sphere,
       and may even have already been detected for a few low-mass objects
       in the form of H~$\alpha$ emission \citep{close14,quanz15,sallum15}.
\end{enumerate}

The next steps will be to extend our analysis to cases of a non-constant EOS to obtain realistic values
for the efficiencies, and to verify the assumption of perfect gas--radiation coupling (the 1-$T$ assumption)
with 2-$T$ radiation transport calculations.
Then, we will couple these efficiency results to formation calculations,
especially in the framework of population synthesis, to make predictions of the post-formation luminosity of gas giants.

Beyond this, due to the generality of our approach, we can easily perform these shock calculations
not only in the context of core accretion but also more generally.
Indeed, these calculations apply also to magnetospheric accretion \citep{koenigl91,lovelace11},
where high-density accretion columns hit the surface of the star;
a similar accretion geometry is a possibility in the context of planet formation (\citealp{katarzy16}; Marleau et al., in prep.).
Also we could easily adapt the parameters (mass, shock radius) to values appropriate
for the flow geometry revealed by global three-dimensional simulations \citep{dangelokley03,tanigawa12,szul16},
where gas falls from high latitudes and shocks on the circumplanetary disk.

\acknowledgements

The authors acknowledge the valuable support of Th.~Henning for this project.
This work has benefitted greatly from discussions with P.~Molli\`{e}re,
\Ae{and we thank also the referee, G.~Chabrier, as well as
A.~Cumming, N.~Turner, W.~Benz, W.~Kley, 
M.~Ikoma, and R.~Pudritz
for discussions and insightful comments.
K.-M.~Ditt\-krist, M.~Schulik,
S.~Ataiee, and A.~Emsenhuber are also thanked
for useful conversations}.
The simulations presented here were performed on the \texttt{ba(t)chelor} cluster
at the MPIA.
G-DM gratefully acknowledges a research fellowship of the International Max-Planck Research School for Astronomy and Cosmic Physics in Heidelberg (IMPRS-HD).
G-DM and CM acknowledge support from the Swiss National Science Foundation under grant BSSGI0\_155816 ``PlanetsInTime''.
Parts of this work have been carried out within the frame of the National Centre for Competence in Research PlanetS supported by the SNSF. 
RK acknowledges financial support within the Emmy Noether research group on
``Accretion Flows and Feedback in Realistic Models of Massive Star Formation''
funded by the German Research Foundation under grant no.~KU 2849/3-1.

\appendix

\section{Relevant parameter space}
 \label{Anh:Schockabschaetzung}
\K{
Here we show what the relevant values of parameter space are,
both in terms of the main formation parameters $\MP$, $\RP$, $\MPunkt$, and $\Lint$,
and of the resulting microphysical parameters $\gamma$, $\mu$, and $\kappa$.
The formation parameter space in which the shock can play a role
covers a few orders of magnitude in each variable,
but this roughly four-dimensional space is by far not uniformly filled.
Moreover, several authors have presented point estimates of the shock temperature
and pressure $\TSch$ and $\Pram$ \citep[e.g.,][]{zhu15},
but this does not give a sense of the range of possible values.
}

\begin{figure}
 \epsscale{0.6}
\plotone{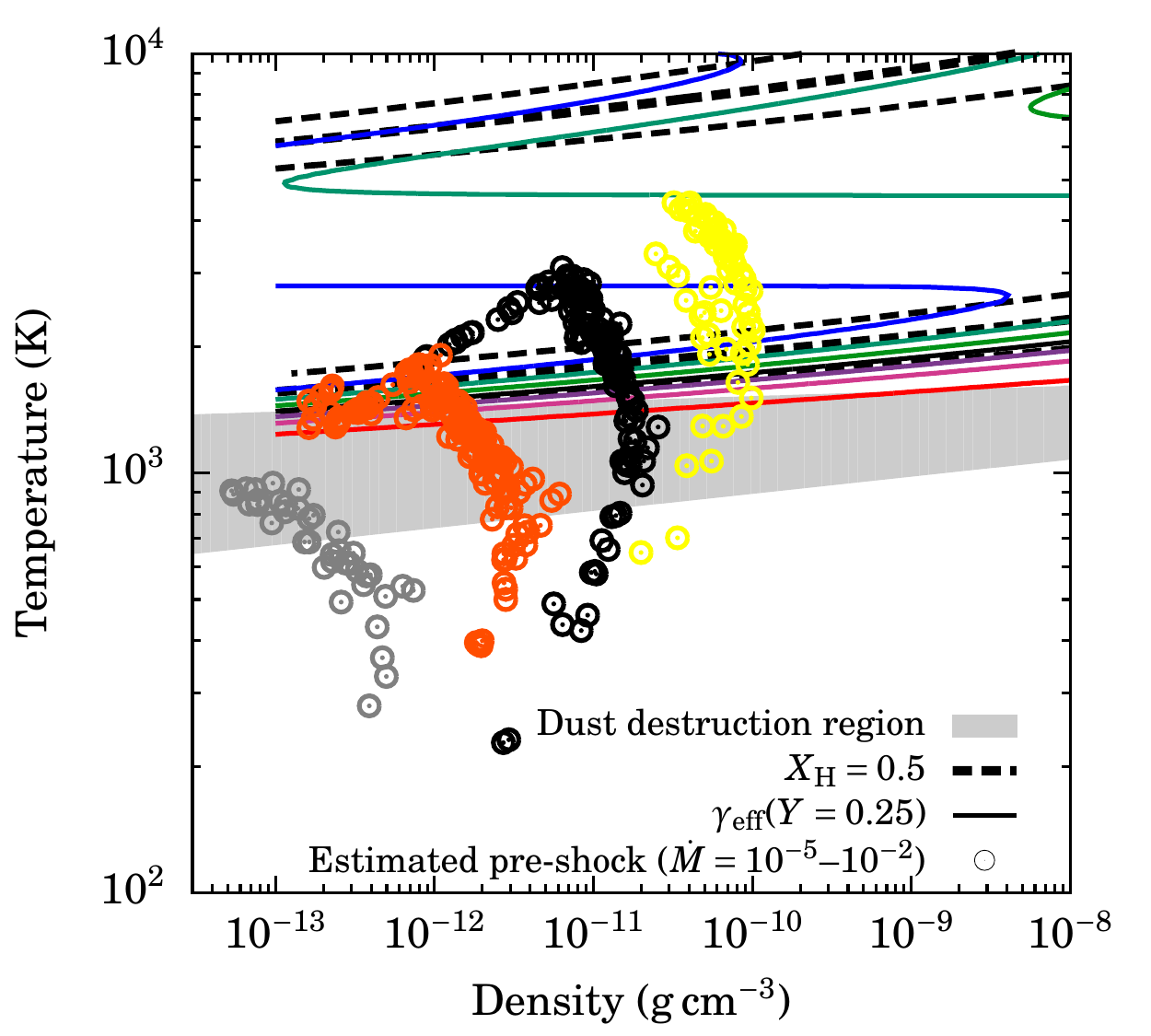}
\caption{
Estimate of the shock temperature
and upstream density
made by assuming $\etaklassisch=1$ (an isothermal shock), $\fred=1$ upstream, and $L(\rSchock^-)\ll L(\rSchock^+)$,
which leads to $4\pi\RP^2 c a T_{\rm shock}^4 \approx G\MP\MPunkt/\RP$ \Aeneu{(see Equation~(\ref{Gl:TSchock_rauh}))},
with $\varrho$ given by the free-fall density \Ae{(Equation~(\ref{Gl:rhoFf}))}.
Shown are contours of $\gamma=1.10$--1.40 in steps of 0.05 (\textit{blue through green to red})
with $\gamma\approx1.4$ from $\approx100$ to 1000~K, as well as
of the ionization or dissociation fraction for hydrogen of $\XHI=0.1$, 0.5, and 0.9 (\textit{black dashed lines}),
and the region of dust destruction in \citetalias{bl94} (\textit{grey band}),
with $\kappa$ of order 1~cm$^2$\,g$^{-1}$ at lower $T$. %
The groups of points are, from left to right,
for $\log\MPunkt/(\ME\,{\rm yr}^{-1})=-5$, $-4$, $-3$, and $-2$.
}
\label{Abb:ungefSchPopSynth_kalt}
\end{figure}

Here we estimate the temperature and density values relevant for the shock
by using the $\MP$, $\RP$, $\MPunkt$, and $\Lint$ values
from the population synthesis of \citet{morda12_I}.
(These data and many more are available
on the Data Analysis Centre for Exoplanets (DACE) platform
at \url{https://dace.unige.ch/evolution/index}.)
Figure~\ref{Abb:ungefSchPopSynth_kalt} shows the lower bound to the shock temperature
for an isothermal shock (Equation~(\ref{Gl:TisothkonstLetakin100}))
using the free-fall velocity (Equation~(\ref{Gl:vFf})),
and the pre-shock density, given by Equation~(\ref{Gl:rhoFf}).
We consider 
$\MP\approx0.2$--$30~\MJ$ and $\MPunkt\approx10^{-4}$--10$^{-2}~\ME\,{\rm yr}^{-1}$,
with $\rSchock\approx1$--30~$\RJ$.
Comparing to the contours of constant $\gamma$
and the rough $\varrho$--$T$ region were dust is destroyed
and the opacity drops from $\sim1$ to $\sim10^{-2}$~cm$^2$\,g$^{-1}$,
one can expect for $\MPunkt\lesssim10^{-5}~\ME\,{\rm yr}^{-1}$
the hydrogen to remain molecular and dust to be only partially destroyed.
At higher accretion rates, however, i.e., for most of the parameter space of interest here,
both dissociation and dust destruction are expected to play a role.

\bibliographystyle{yahapj}

\providecommand\natexlab[1]{#1}

\end{document}